\documentclass[epj]{webofc}
\usepackage[varg]{txfonts}   
\woctitle{Mathematical Modeling and Computational Physics 2015}

\usepackage[english]{babel}

\def\mn{m_N}

\def\qp{{q'}}

\def\mn{m_N}

\def\qz{q_0}
\def\qzp{q_0'}

\def\sq{\sqrt{s}}

\def\bq{{\bf q}}
\def\bqp{{\bf \qp}}

\begin{document}
\title{Relativistic three-nucleon calculations within the Bethe-Salpeter approach}
%
%

\author{S.G. Bondarenko\inst{1}\fnsep\thanks{\email{bondarenko@jinr.ru}} \and
        V.V. Burov\inst{1}\fnsep\thanks{\email{burov@theor.jinr.ru}} \and
        S.A. Yurev\inst{1,2}\fnsep\thanks{\email{yurev@jinr.ru}}
}

\institute{Joint Institute for Nuclear Research, Dubna, 141980, Russia 
\and
           Far Eastern Federal University, Vladivostok, 690950, Russia 
          }

\abstract{%
The relativistic properties of the three-nucleon system are investigated
using the Faddeev equations within the Bethe-Salpeter approach.
The nucleon-nucleon interaction is chosen in a separable form.
The Gauss quadrature method is used to calculate the integrals.
The system of the integral equations are solving by iterations method.
The binding energy and the partial-wave amplitudes ($^1S_0$ and $^3S_1$)
of the triton are found.
}
\maketitle
\section{Introduction}
\label{intro}
Three-body calculations in nuclear physics are of great interest used describing
three-nucleon bound states ($^3He$, $T$), processes of elastic, inelastic and deep
inelastic scattering the leptons off light nuclei and also the hadron-deuteron reactions
(for example, $pd \to pd$, $pd \to ppn$). Study of the nuclei $^3He$ and $T$ is also interesting
because it allows us to investigate further (in addition to the case of the deuteron) evolution
of the bound nucleon thereby contributing to the explanation of so-called EMC-effect.
In quantum mechanics the Faddeev equations are commonly used
to describe the three-particle systems. The main feature of Faddeev equations
is that all particles interact through a pair potential.

However at the high momentum transfer relativistic effects should be taken into account.
The Bethe-Salpeter (BS)~\cite{Salpeter:1951sz} equation
is one of the most consistent approaches to
describe the NN interaction. In this formalism, one has to deal with a
system of nontrivial integral equations for both the NN scattered
states and the bound state -- the deuteron. To solve a system of
integral equations, it is convenient to use a separable
{\it Ansatz}~\cite{Bondarenko:2002zz} for the interaction kernel in the
BS equation. In this case, one can transform integral equations
into a system of algebraic linear ones which is easy to solve.
Parameters of the interaction kernel are found from an analysis
of the phase shifts and inelasticity, low-energy
parameters and deuteron properties (binding energy, moments, etc.).

The relativistic three-particle systems are described by the Faddeev equations
within the BS approach - so called Bethe-Salpeter-Faddeev equations.
In the paper the three-nucleon nuclei is studied with the simplest Yamaguchi
separable potential~\cite{yam}. All nucleons have equal masses and the scalar
propagators instead of spinor ones are used for simplicity.
The spin-isospin structure of the nucleons is taken into account
by using the so-called recoupling-coefficient matrix. The work mainly follows the ideas
of the article~\cite{Ref1}.

The paper is organized as following: in Sec.~2 the two-particle problem is considered,
in Sec.~3 - three-particle equations. In Sec.~4 the calculations and results are given.
The summary is in the Sec.~5.

\section{Two particle case}
\label{sec-1}

Since the formalism of the Faddeev equations is based on the properties of the pair nucleon-nucleon
interaction only some conclusions of two-body problem are given here.

The Bethe-Salpeter equation for the relativistic two-particle system is taken in the following form:
\begin{equation}
 T(p,p';s) =  V(p,p') +  \frac{i}{4\pi^3}\int d^4k\, V(p,k)\, G(k;s)\, T(k,p';s)
\label{eqn1}
\end{equation}
where $T(p,p';s)$ is the two-particle $T$ matrix and $V(p,p')$ - kernel (potential) of the nucleon-nucleon
interaction. The free two-particle Green function $G(k;s)$ is expressed, for simplicity, thought the scalar
propagator of the nucleons
\begin{equation}
G^{-1}(k;s) = \big[(P/2 + k)^2 - \mn^2 + i\epsilon \big] \big[(P/2 - k)^2 - \mn^2+ i\epsilon\big].
\label{eqn2}
\end{equation}

To solve equation~(\ref{eqn1}) the separable {\it Ansatz} for the nucleon-nucleon potential $V(p,p')$ is used
(rank-one)
\begin{equation}
V(p_0,p,p_0',p') = \lambda g(p_0,p) g(p_0',p').
\label{eqn3}
\end{equation}
In this case the two-particle $T$ matrix has the following simple form:
\begin{equation}
T(p_0,p,p_0',p';s) = \tau(s) g(p_0,p) g(p_0',p')
\label{eqn4}
\end{equation}
where
\begin{equation}
[\tau(s)]^{-1} = {\lambda}^{-1} - \frac{i}{4\pi^3}\int_{-\infty}^{\infty} dk^0 \int_0^\infty k^2\, dk\, g^2(k^0,k)\,
G(k^0,k;s).
\end{equation}

As a simplest assumption the relativistic {\it Yamaguchi}-type form factor $g_Y(p_0,p)$ is used
\begin{equation}
g_Y(p_0,p) = \frac{1}{-p^2_0 + p^2 + \beta^2},
\label{eqn5}
\end{equation}
with parameters  $\beta$ and $\lambda$ are chosen to describe experimental data (deuteron
binding energy, low-energy scattering parameter and phase shifts).
The values of the parameters are given in Table~\ref{tab1}.

\begin{table}[htb]
\centering
\caption{Parameters for $\tau(s)$ and  $g_Y(p_0,p)$ for S-states}
\label{tab1}  
\begin{tabular}{ccc}
\hline
Parameter & $_{}^1S_0$ & $_{}^3S_1$ \\ \hline
$\lambda$ (GeV$^{4}$) & -1.12087 & -3.15480 \\
$\beta$ (GeV) & 0.287614 & 0.279731 \\ \hline
\end{tabular}
\end{table}

\section{Three-particle case}
\label{3particle}
The relativistic three-particle system can be described by the Bethe-Salpeter-Faddeev equations
\begin{eqnarray}
\Biggl[
\begin{array}{c}
T^{(1)}\\
T^{(2)}\\
T^{(3)}
\end{array}
\Biggr]
=
\Biggl[
\begin{array}{c}
T_{1}\\
T_{2}\\
T_{3}
\end{array}
\Biggr]
-
\Biggl[
\begin{array}{ccc}
0      & T_1G_1 & T_1G_1 \\
T_2G_2 & 0      & T_2G_2 \\
T_3G_3 & T_3G_3 & 0
\end{array}
\Biggr]
\Biggl[
\begin{array}{c}
T^{(1)}\\
T^{(2)}\\
T^{(3)}
\end{array}
\Biggr]
\label{eqn6}
\end{eqnarray}
where full matrix $T=\sum_{i=1}^3T^{(i)}$, $G_i$ is the two-particle ($j$ and $n$)
Green function ($ijn$ is cyclic permutation of (1,2,3)):
\begin{eqnarray}
G_i(k_j,k_n) = 1/(k_j^2-\mn^2+i\epsilon)/(k_n^2-\mn^2+i\epsilon),
\label{eqn7}
\end{eqnarray}
and $T_i$ is the two-particle $T$ matrix which can be written as following:
\begin{eqnarray}
T_i(k_1,k_2,k_3;k_1',k_2',k_3') = (2\pi)^4 \delta^{(4)}(K_i-K_i') T_i(k_j,k_n;k_j',k_n').
\label{eqn8}
\end{eqnarray}

For the system of equal-mass particles the Jacobi momenta can be written in the following form:
\begin{eqnarray}
p_i = \frac12 (k_j-k_n),\ q_i = \frac13 K - k_i,\ K=k_1+k_2+k_3.
\label{eqn9}
\end{eqnarray}
Using expressions~(\ref{eqn9}) the equation~(\ref{eqn6}) can be rewritten as
\begin{eqnarray}
&&T^{(i)}(p_i,q_i;p_i',q_i';s) = (2\pi)^4 \delta^{(4)}(q_i-q_i') T_i(p_i;p_i';s)
\label{eqn10}\\
&&-i\int\frac{d p_i''} {(2\pi)^4}
T_i(p_i;p_i'';s) G_i(k_j'',k_n'') [
T^{(j)}(p_j'',q_i'';p_i',q_i';s) + T^{(n)}(p_i'',q_i'';p_i',q_i';s)
]
\nonumber
\end{eqnarray}

It is suitable to introduce the amplitude $\Psi^{(i)}(p_i,q_i;s)$
for the bound state of the three particles in the following form:
\begin{eqnarray}
\Psi^{(i)}(p_i,q_i;s) = \langle p_i,q_i | T^{(i)} | M_B \rangle \equiv 
\Psi_{L M}(p,q;s),
\label{eqn11}
\end{eqnarray}
where $M_B = \sqrt{s} = 3\mn-E_{B}$ is the mass of the bound state (triton)
and $s=K^2$ is the total momentum squared. For the equal-mass case all
$\Psi^{(i)}$ functions are equal to each other.

The total orbital angular momenta of the triton can be presented as {\boldmath{${L} = {l} + {\lambda}$}},
where {\boldmath $l$} is the angular momentum corresponding to nucleon pair with relative impulse {\boldmath $p$}
and {\boldmath $\lambda$} is the angular momentum corresponding to relative impulse {\boldmath $q$}.

To separate the angular dependence the amplitude can be written in the following form:
\begin{eqnarray}
\Psi_{L M}(p,q;s) = \sum_{a\lambda} \Psi^{(a)}_{\lambda L}(p_0,|{\bf p}|,q_0,|{\bf q}|;s) {\cal Y}^{(a)}_{\lambda LM}({\hat {\bf p}},{\hat {\bf q}}),
\label{eqn12}
\end{eqnarray}
with the angular part
\begin{eqnarray}
{\cal Y}^{(a)}_{\lambda LM}({\hat {\bf p}},{\hat {\bf q}}) =
\sum_{m\mu} C_{lm\lambda\mu}^{LM} Y_{lm}({\hat {\bf p}}) Y_{\lambda\mu}({\hat {\bf q}}),
\label{eqn12a}
\end{eqnarray}
where the two-nucleon state with spin $s$, angular moment $l$ and total momentum $j$  $(a\equiv{^{2s+1}l_j})$
is introduced, $C$ is the Clebsch-Gordan coefficients and $Y$ is spherical harmonics.

If one consider the rank-one separable two-nucleon interaction the amplitude
$\Psi_{\lambda L}^{(a)}$ can be written as
\begin{eqnarray}
\Psi^{(a)}_{\lambda L}(p_0,p,q_0,q;s) = g^{(a)}(p_0,p) \tau^{(a)}(s) \Phi^{(a)}_{\lambda L}(q_0,q;s),
\label{eqn12b}
\end{eqnarray}
where function $\Phi^{(a)}_{\lambda L}$ satisfies the following integral equation:
\begin{eqnarray}
\Phi^{(a)}_{\lambda L}(q_0,q;s) =  \frac{i}{4\pi^3} \sum_{a'\lambda'} \int_{-\infty}^{\infty} dq_0'\int_{0}^{\infty}q^{'2}dq'\,
Z^{(aa')}_{\lambda\lambda'}(q_0,q;q_0',q';s)
\\
\frac{\tau^{(a')}[(\frac{2}{3}\sqrt s+q_0')^2-q'^{2}]}{(\frac{1}{3}\sqrt s-q_0')^2-q^{'2}-m^2+i\epsilon}
\Phi^{(a')}_{\lambda'}(q_0',q';s),
\nonumber
\end{eqnarray}
with
\begin{eqnarray}
Z^{(aa')}_{\lambda\lambda'}(q_0,q;q_0',q';s) = C_{(aa')}\int d\cos\vartheta_{qq'}
K^{(aa')}_{\lambda\lambda' L}(q,q',\cos\vartheta_{qq'})\\
\frac{
g^{(a)}(-\frac{1}{2}q_0-q'_0,|{\bf q}/2+{\bf q'}|)
g^{(a')}(q_0+\frac{1}{2}q'_0,|{\bf q}+{\bf q'}/2|)}
{(\frac{1}{3}\sqrt s +q_0+q'_0)^2-({\bf q}+{\bf q'})^2-\mn^2+i\epsilon}
\nonumber
\end{eqnarray}
and
\begin{eqnarray}
K^{(aa')}_{\lambda\lambda' L}(q,q',\cos\vartheta_{qq'}) =
(4\pi)^{3/2}\frac{\sqrt{2\lambda + 1}}{2L+1}\\
\sum_{mm'}
C^{L m}_{l m \lambda 0}
C^{L m}_{l' m' \lambda' m-m'}
{Y^{*}}_{l m}(\vartheta,0) 
{Y}_{l' m'}(\vartheta',0),
{Y}_{\lambda' m-m'}(\vartheta_{qq'},0)
\nonumber
\end{eqnarray}
where
$$\cos\vartheta=(\frac{q}{2} + q'\cos\vartheta_{qq'})/|\frac{\bq}{2}+\bqp|,\quad\quad\cos\vartheta'=(q+\frac{q'}{2}\cos\vartheta_{qq'})/|\bq+\frac{\bqp}{2}|$$
and
$C_{(aa')}$ is the spin-isospin recoupling-coefficient matrix.

Considering the ground state of the three-particles state $L=0$
and assuming the internal orbital angular momenta are equal to zero
$l = \lambda = 0$ one should take into into account only two states:
$a = (^1S_0,^3S_1)$.
In this case function $K^{00}_{000} = 1$ and system of integral equations
reads
\begin{eqnarray}
\Phi^{(a)} (\qz,q;s) = \frac{i}{4\pi^3} \sum_{a'} \int d\qzp \int{\qp}^2d\qp\,
Z^{(aa')}(\qz,q,\qzp,\qp;s)
\label{eqn13}\\
\frac{\tau^{(a')}[(\frac23 \sq + \qzp)^2-\qp^2]}{(\frac13 \sq - \qzp)^2-\qp^2-\mn^2+i\epsilon}
\Phi^{(a')}(\qzp,\qp;s).
\nonumber
\end{eqnarray}
Here the so-called effective energy-dependent potential $Z$ is 
\begin{eqnarray}
Z^{(aa')}(\qz,q,\qzp,\qp;s) = C_{(aa')} \int_{-1}^{1}d(cos\vartheta_{qq'})
\label{eqn14}\\
\frac{g^{(a)}(- \frac12 \qz - \qzp,|- \frac12 \bq - \bqp|) 
g^{(a')}(\qz + \frac12 \qzp,|\bq + \frac12 \bqp|)}
{(\frac13 \sq + \qz + \qzp)^2-(\bq + \bqp)^2-\mn^2+i\epsilon},
\nonumber
\end{eqnarray}
with
\begin{eqnarray}
C_{(aa')}
=
\Biggl[
\begin{array}{cc}
\frac{1}{4}     & -\frac{3}{4}   \\
-\frac{3}{4}  & \frac{1}{4}        
\end{array}
\Biggr]
\label{eqn15}
\end{eqnarray}

The system of integral equations~(\ref{eqn13}-\ref{eqn14}) has the number of singularities,
however in the case of the bound three-particle system ($\sqrt s < 3\mn$) all this singularities
do not cross the path of integration on $q_0$ and thus do not affect to the Wick-rotation procedure
$q_0 \rightarrow  iq_4$.

The system of Eqs.~(\ref{eqn13}-\ref{eqn14}) after the Wick-rotation procedure
is well analytically defined and can be
solved by different numerical methods. One of them is discussed in the next section.

\section{Solution and results}
\label{results}

In order to solve the system of integral equations the iterations method is used.
The mappings for the variable of integration on $q$ $[0,\infty)$ and $q_4$ $(-\infty,\infty)$ to
$[-1,1]$ interval are introduced. 

The binding energy of the three-nucleon system is satisfied the following condition (see details in~\cite{Method}):
\begin{eqnarray}
\lim_{n \to \infty}\frac{\Phi_n(s)}{\Phi_{n-1}(s)}\Big|_{s=M_B^2} = 1 
\end{eqnarray}
where n is number of iteration.

Result of calculations for the binding energy is $E_B$ = 10.99 MeV which should be compared
to the experimental value 8.48 MeV. The difference can be explained
by the simplicity of the using separable kernel (rank-one) of the nucleon-nucleon interaction.

\begin{figure}[htb]
\centering
\begin{minipage}{\textwidth}
\includegraphics[width=0.5\textwidth]{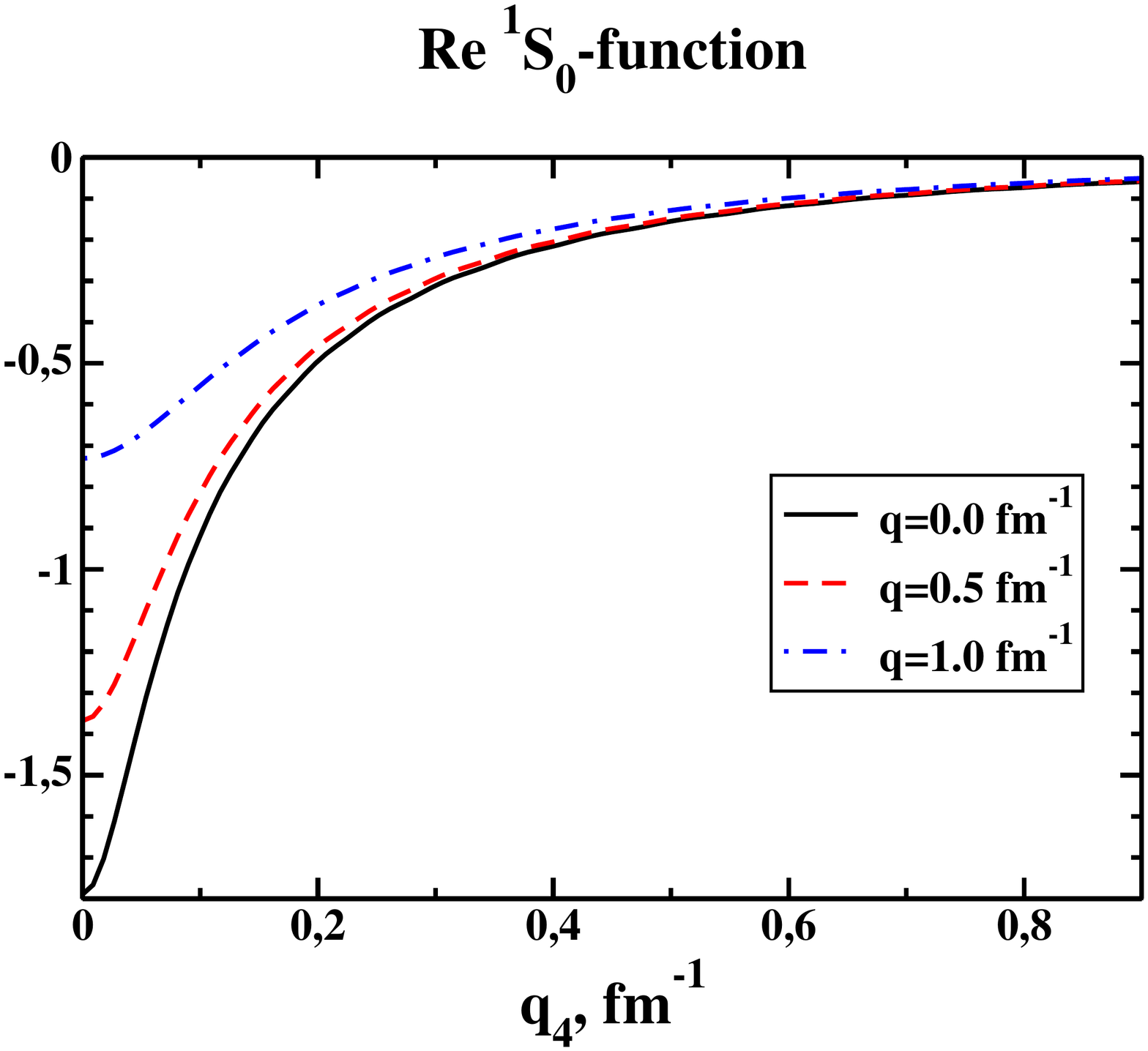}
\hskip 5mm
\includegraphics[width=0.5\textwidth]{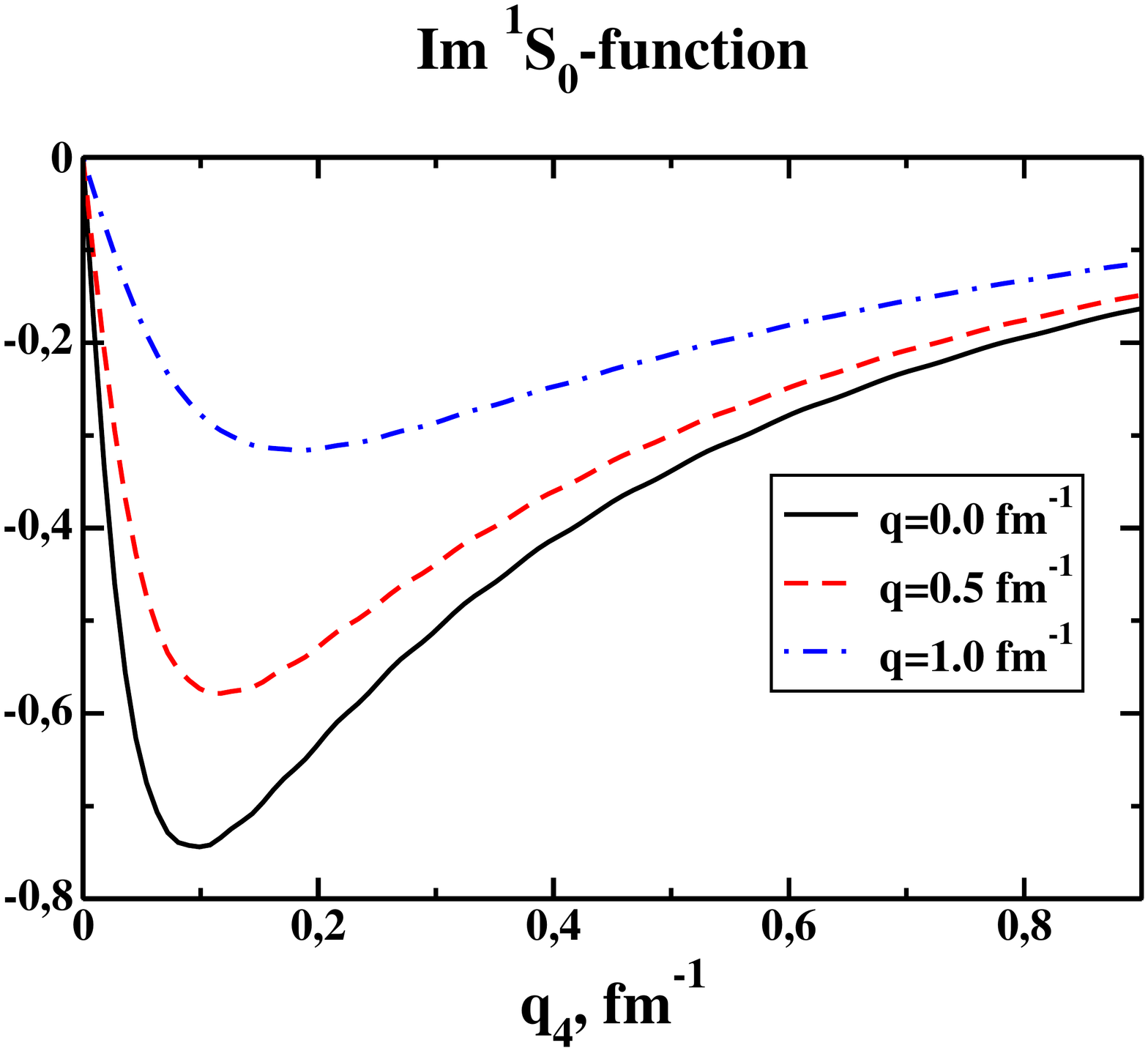}
\end{minipage}
\caption{Real (left) and imaginary (right) parts of the $^1S_0$ partial-wave state
as a functions of $q_4$ at different $q$ values: solid black line - for $q = 0$ Fm$^{-1}$,
dashed red line - for $q = 0.5$ Fm$^{-1}$ and dotted-dashed blue line - for $q = 1$ Fm$^{-1}$}
\label{f1}     
\end{figure}

\begin{figure}[htb]
\centering
\begin{minipage}{\textwidth}
\includegraphics[width=0.5\textwidth]{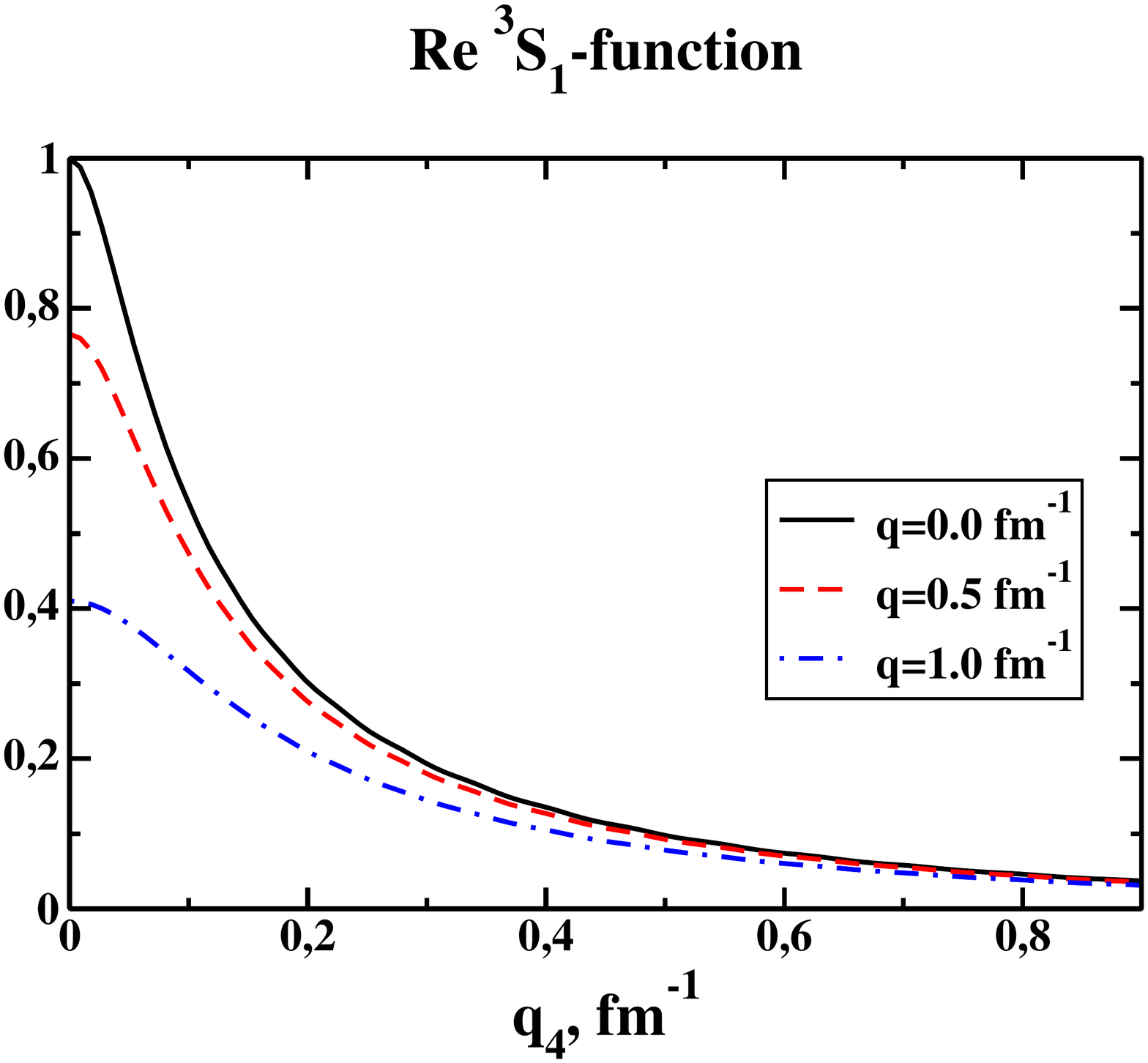}
\hskip 5mm
\includegraphics[width=0.5\textwidth]{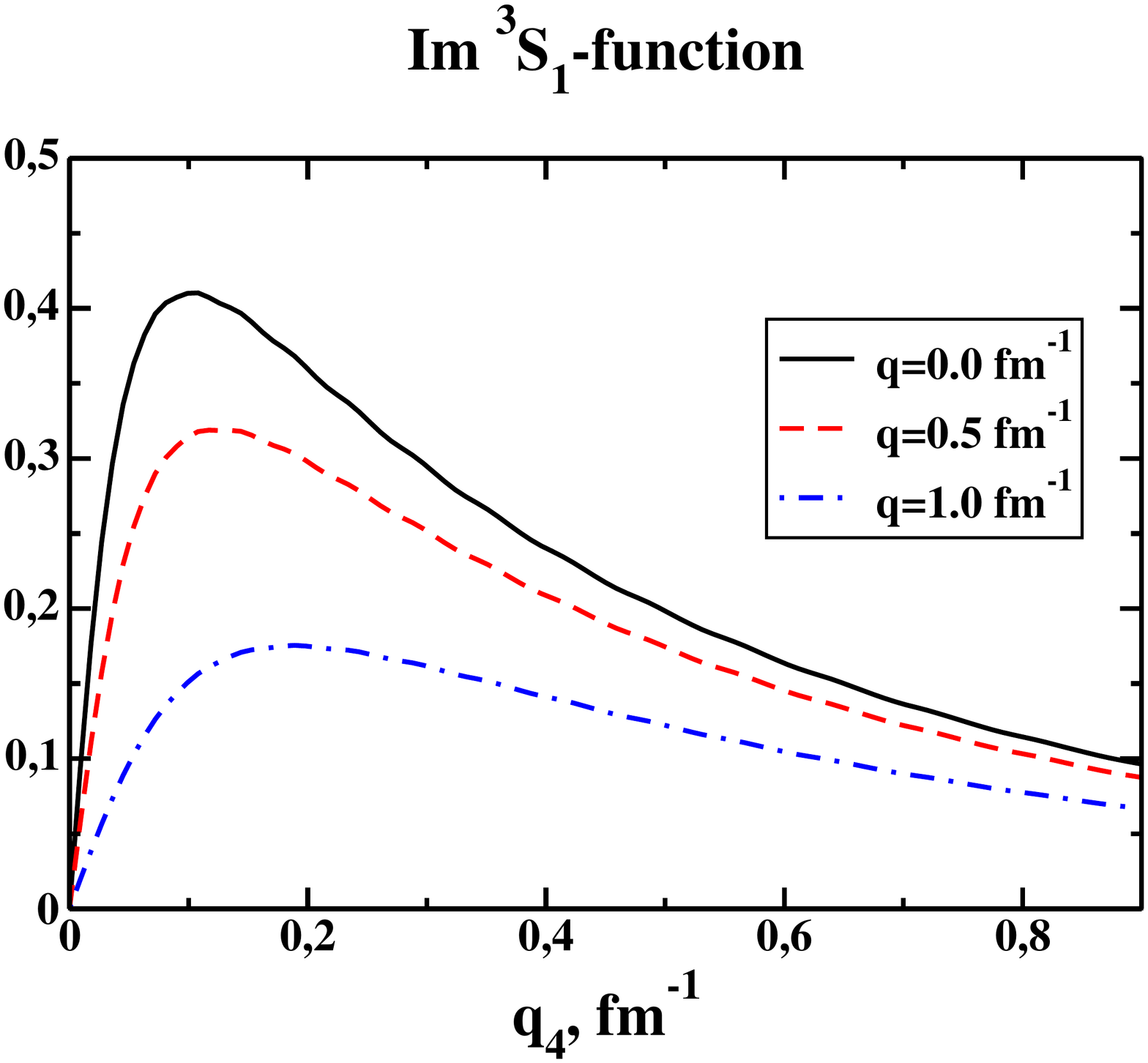}
\end{minipage}
\caption{Real (left) and imaginary (right) parts of the $^3S_1$ partial-wave state
as a functions of $q_4$ at different $q$ values: solid black line - for $q = 0$ Fm$^{-1}$,
dashed red line - for $q = 0.5$ Fm$^{-1}$ and dotted-dashed blue line - for $q = 1$ Fm$^{-1}$}
\label{f2}     
\end{figure}

\begin{figure}[htb]
\centering
\begin{minipage}{\textwidth}
\includegraphics[width=0.5\textwidth]{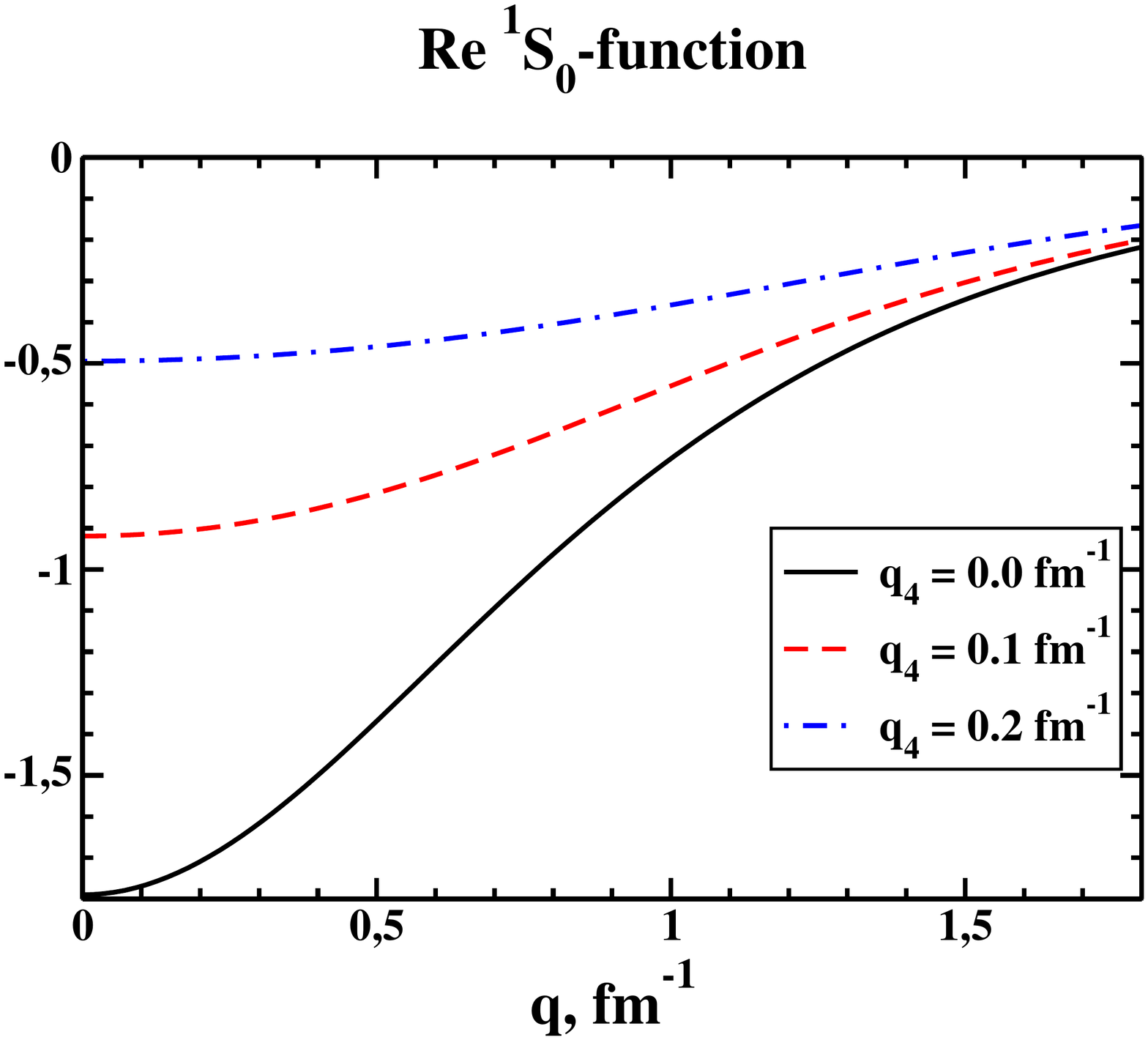}
\hskip 5mm
\includegraphics[width=0.5\textwidth]{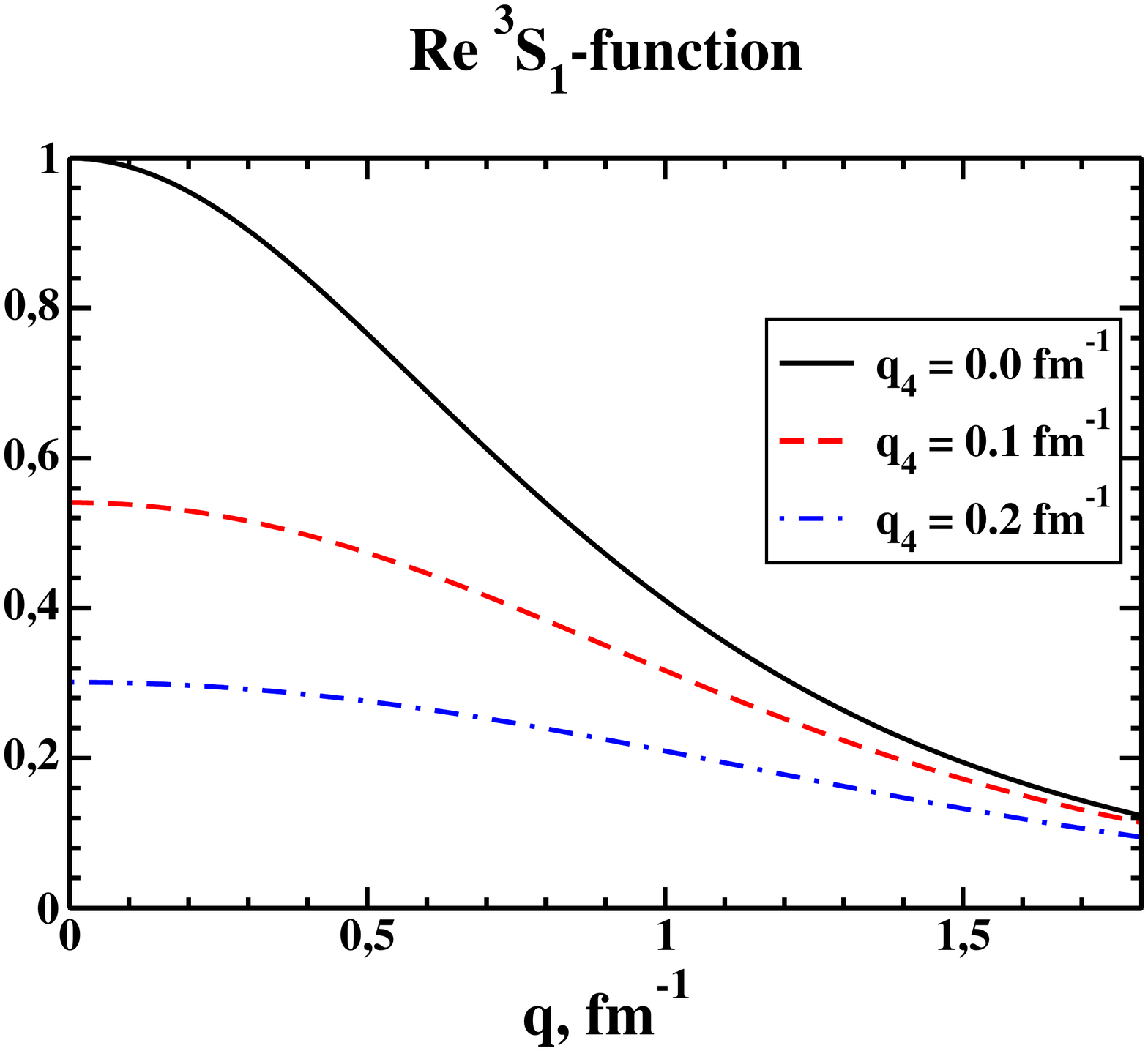}
\end{minipage}
\caption{Real parts of the $^1S_0$(left) and $^3S_1$(right) partial-wave state
as a functions of $q$ at different $q_4$ values: solid black line - for $q_4 = 0$ Fm$^{-1}$,
dashed red line - for $q_4 = 0.1$ Fm$^{-1}$ and dotted-dashed blue line - for $q_4 = 0.2$ Fm$^{-1}$}
\label{f3}     
\end{figure}

The obtained partial-wave amplitudes are shown in the Figs.~(\ref{f1}-\ref{f3}).

\section{Summary}
\label{summary}

In the paper three-body system is investigated by using Bethe-Salpeter-Faddeev equations.
The nucleon-nucleon interaction is taken in the rank-one separable form.
The relativistic generalization of Yamaguchi-type functions are chosen in the calculations.
The parameters of the nucleon-nucleon potential for the $^1S_0$ and $^3S_1$ partial-wave states
reproduce the low-energy scattering parameter and deuteron properties as well as phase shifts up to
the laboratory energy 100-120 MeV.
The BSF integral equations are solved by using iterations method.
The binding energy of the triton and amplitudes of the $^1S_0$ and $^3S_1$
partial-wave states of the triton are found.

The large overestimation of the binding energy of the triton is found. To improve results the rank
of the separable kernel should be increased. Also another partial-wave states, such as $P$- and $D$-states,
and spinor propagators for nucleons should be taken into account.

\section*{Acknowledgments}
One of the authors (S.G. Bondarenko) thanks organizers of the
International Conference ``Mathematical Modeling and Computational Physics, 2015''
(Star\'a Lesn\'a, Slovakia, July 13 — 17, 2015)
and personally Professor Michal Hnati\v{c} for invitation, support and hospitality.


\begin{thebibliography}{}
%
%
\bibitem{Salpeter:1951sz}
E.E. Salpeter, H.A. Bethe, Phys. Rev. {\bf 84}, 1232 (1951).

\bibitem{Bondarenko:2002zz}
S.G. Bondarenko, V.V. Burov, A.V. Molochkov, G.I. Smirnov, H. Toki,
Prog. Part. Nucl. Phys. {\bf 48}, 449 (2002).







\bibitem{yam} Y.~Yamaguchi, Phys. Rev. {\bfseries 95}, 1628 (1954);
Y.~Yamaguchi, Y.~Yamaguchi, Phys. Rev. {\bfseries 95}, 1635 (1954).

\bibitem{Ref1} G. Rupp and J. A. Tjon,
Phys. Rev. C {\bf 37}, 1729 (1988).


\bibitem{Method} R. A. Malfliet and J. A. Tjon, Nucl. Phys. {\bf A127},161-168 (1969)

\end{thebibliography}
\end{document}